\input{epsf}
\documentstyle[preprint,aps]{revtex}
\begin{document}

\title{Quasinormal modes of nearly extreme Reissner-Nordstr\"om black holes
\footnote{Preprint number: TIT/HEP-339/COSMO-76 and WUGRAV96-7}}

\author{Nils Andersson $\dag$ and Hisashi Onozawa $\ddag$}

\address{$\dag$ Department of Physics, Washington University, 
St Louis MO 63130, USA}

\address{$\ddag$ Department of Physics, Tokyo Institute of Technology, Oh-okayama, Meguro, Tokyo 152, Japan}

\maketitle

\begin{abstract}
We present detailed calculations of the quasinormal modes of
Reissner-Nordstr\"om black holes. While the first few, slowly damped,
modes depend on the charge of the black hole 
in a relatively simple way, we find that the
rapidly damped modes show several peculiar features. The higher modes
generally spiral into the value for the extreme black hole as the
charge increases. We also discuss the possible existence of a purely
imaginary mode for the Schwarzschild black hole:  Our data suggest that
there is a quasinormal mode that limits to $\omega M = -2i$ as $Q\to
0$. 
\end{abstract}


\section{Perturbing the Reissner-Nordstr\"om black hole}

Quasinormal modes of black holes have been studied ever since the
seminal work of Vishveshwara \cite{vishu} and Press \cite{press} in the
early 1970s. It soon became evident that exponentially damped
mode-oscillations will dominate most processes involving perturbed
black holes (see \cite{nbook} for references).  This means that the
quasinormal modes provide a unique opportunity to identify a black
hole, a possibility that hopefully will become reality when large-scale
laser-interferometric detectors for gravitational waves come into
operation in the near future \cite{ligo}.  In order to extract as much
information as possible from a gravitational-wave signal it is
important that we understand exactly how the quasinormal modes depend
on the parameters of the black hole.

The parameters of main astrophysical importance are the black holes
mass and angular momentum. That is, the Kerr solution is the most
relevant one from an astrophysical point of view.  The solution that
describes an electrically charged, nonrotating black hole --- due to
Reissner and Nordstr\"om  --- is of less direct importance because it
seems unlikely that black holes with a considerable charge will exist
in the Universe.  Nevertheless, the Reissner-Nordstr\"om solution has
several interesting features that warrant a closer inspection. The
most intriguing one concerns the possible conversion of electromagnetic
energy into gravitational energy and vice versa:  In a charged
environment an electromagnetic wave will inevitably give rise to
gravitational waves. The Reissner-Nordstr\"om metric provides the
simplest framework for studies of this effect.

For this and several other reasons there has been a number of studies
of perturbed Reissner-Nordstr\"om black holes. The equations governing
a weak (massless) field in the geometry of an electrically charged
black hole were first derived by Zerilli \cite{zerilli} and Moncrief
\cite{moncrief}.  Basically, these equations can be i) split into axial
and polar perturbations (also known as odd and even parity,
respectively) and then ii) reduced to two decoupled wave equations in
each case. The final  wave equations describe two variables $\Psi_1$
and $\Psi_2$, from which all components of the electromagnetic field
and the perturbed metric can be reconstructed. In general, both
$\Psi_1$ and $\Psi_2$ correspond to a combination of electromagnetic
waves and gravitational ones, but in the limiting case of an uncharged
black hole the two functions reduce to pure electromagnetic and
gravitational waves, respectively.

Originally, the perturbation equations were used to study the stability
of the metric \cite{zerilli,moncrief}. Later the equations were used to
investigate the already mentioned conversion of electromagnetic energy
 into gravitational \cite{johnston,olson,gunter}. There has also been
several studies of quasinormal modes for Reissner-Nordstr\"om black
holes \cite{kokkotas,leaver,andersson}.  But although the methods used
in those studies provide
 accurate numerical results there is still need for more information.
Specifically, one would like to know what happens to the quasinormal
modes as the black hole becomes extremely charged. There are two parts
to this problem, that seemingly must be studied separately. In the
typical case each of the two horizons, at $r_\pm = M\pm\sqrt{M^2-Q^2}$
(where $M$ is the mass and $Q\le M$ the charge of the hole), correspond
to a second order singularity in the linearized equations. But in the
extreme case these two singularities merge (at $r=M$) into a single one
of fourth order. This means that the methods that have been devised to
determine the entire spectrum of quasinormal modes break down in the extreme limit \cite{leaver,andersson}. 

Even though the extremely charged
case is only of theoretical interest it is worth considering. One
reason (apart from natural curiosity) is the fact that the existence of
two coalescing horizons is a prominent feature also of rapidly rotating
black holes. Thus one may hope that a study of the mode-behaviour as
$Q\to M$ can lead to some insights also for the rotating case. Or at
least that methods that prove reliable for high charge can be adapted
to the Kerr case.  Recently Leaver's continued fraction method was
amended in such a way that it could be applied to the case of extreme charge
\cite{onozawa}.  Although it seems plausible that other methods, such
as  the
numerical integration of Andersson \cite{andersson}, can be extended in
a similar way there has been no attempts to do this.

Anyway, at present the main question concerns the reliability of the
various approaches for nearly extreme black holes. It is clear that the
methods in \cite{leaver,andersson} will break down for $Q=M$,
and also that the method of Onozawa {\em et al.} \cite{onozawa} can  be
used only for the extreme case.  Thus, we do not yet know to what
extent the available results for high charge (and perhaps also rapid
rotation in the case of  Kerr \cite{leaverprs}) can be trusted. The
work presented in this paper was motivated by a desire to obtain a
better understanding of this issue. We also wanted to unveil the
detailed behaviour of the quasinormal modes as the charge of the black
hole was increased. Specifically, previous evidence (see Figure 1 in
\cite{leaver} and Figures 2 and 3 in \cite{andersson}) indicate that
this behaviour is somewhat peculiar for the higher overtones of the
black hole.  As we will show in the following section, the behaviour of
the highly damped Reissner-Nordstr\"om quasinormal modes is, indeed,
very strange.
 
\section{Presenting the results}

\subsection{The numerical work}

Since the equations that describe a perturbed Reissner-Nordstr\"om
black hole are available in the literature (most notably in
Chandrasekhar's exhaustive book \cite{cbook}) we will not list them
here. For the present discussion it is
sufficient to know that the  equations take the general form
\begin{equation}
{d^2\Psi \over dr_\ast^2} + [ \omega^2 - V(r) ] \Psi = 0 \ ,
\label{rweq}\end{equation}
where we have assumed that the time-dependence of the perturbation is
$e^{-i\omega t}$. The tortoise coordinate, that is defined by
\begin{equation}
{d\over dr_\ast} = \left( 1 - {2M\over r} + {Q^2 \over r^2} \right) {d\over dr} \ ,
\label{tortoise}\end{equation}
has the effect that the event horizon of the black hole is 
``pushed away'' to 
$r_\ast = -\infty$.

The function $\Psi$ can be either an axial (odd parity) or polar (even
parity) perturbation of the black hole. Also, for each case there are
two different functions, $\Psi_1$ and $\Psi_2$, that correspond to pure
electromagnetic and gravitational waves in the Schwarzschild limit. In
total there are thus four equations, all with slightly different
effective potentials. For obvious reasons this difference in the
potentials leads to somewhat different quasinormal modes for $\Psi_1$
and $\Psi_2$. Analogously, one might expect the quasinormal modes for
axial and polar perturbations to be different. But, as pointed out by
Chandrasekhar \cite{cbook}, the mathematical theory of black holes is
an intricately entangled web where many quantities are related (often in a
surprising way). Thus it turns out that the quasinormal modes for axial
and polar perturbations are identical. The underlying reason is that
two effective potentials can, although different, contain much the same
physical information.  This means that it is sufficient to restrict a
study of the Reissner-Nordstr\"om problem to either the axial or the
polar case. Once $\Psi_1$ and $\Psi_2$ are found for one kind of
perturbation the corresponding solutions for the other case are easily
generated \cite{cbook}.

The quasinormal modes of a black hole correspond to solutions to 
Eq.~(\ref{rweq}) that satisfy the causal condition that no information
should leak out through the event horizon of the black hole, and at the
same time correspond to purely outgoing waves at spatial infinity. This
means that a typical mode-solution will behave like
\begin{equation}
\Psi \sim \left\{ \begin{array}{ll}  e^{i\omega r_\ast} \ , \mbox{ as } r_\ast \to +\infty \ , \\
 e^{-i\omega r_\ast} \ , \mbox{ as } r_\ast \to -\infty \ .
\end{array} \right.
\label{asymp}\end{equation}
Since one would expect the black hole to be stable against a small
perturbation the mode-frequencies should be complex. This implies that
an identification of quasinormal modes is non-trivial. In order
for a mode-solution to be damped with time at a fixed $r_\ast$ the
frequency $\omega$ must have a negative imaginary part. But then the
corresponding solution will diverge  both at the event horizon and
spatial infinity [cf. Eq.~(\ref{asymp})] at a fixed time.  Several methods
have been devised to deal with this difficulty. Most notable are
Leaver's continued fraction approach \cite{leaverprs} and Andersson's
complex-coordinate integration method \cite{anderssonprs}. These two
methods provide highly accurate numerical results also for rapidly
damped modes.  Furthermore, both methods have been used to study the
first few quasinormal modes for Reissner-Nordstr\"om black holes.

As already mentioned in the previous section, it is clear that all
existent methods will fail for a nearly extreme black hole. The reason
is that the two horizons of the black hole merge as $Q\to M$, and this
changes the singularity structure of the problem (this is easy to see
if Eq.~(\ref{rweq}) is expanded in such a way that all derivatives are
taken with respect to $r$ rather than $r_\ast$). Thus, the extreme case
must be considered separately.  This was recently done by Onozawa {\em
et al.} \cite{onozawa}.  As a test of the reliability of that study,
and also to provide a better understanding of the behaviour of the
quasinormal modes for highly charged black holes we decided to make an
exhaustive study of the problem. We made detailed calculations [using
both the continued fraction method \cite{leaver} and numerical
integration \cite{andersson} to make sure that the results were
reliable] for the first nine dipole ($\ell=1$) modes of $\Psi_1$ and
the first nine quadrupole ($\ell=2$) modes of $\Psi_2$. These are the
lowest radiating multipoles for each case. In general one finds, just
as for Schwarzschild black holes \cite{nbook}, that the mode-behaviour
is similar for higher values of $\ell$.  
The results of this investigation are shown in Figs.~\ref{fig1} and \ref{fig2}. We now proceed to discuss them in more detail.

\subsection{Slowly damped modes}

\begin{figure}
\centerline{\epsfysize=450pt \epsfbox{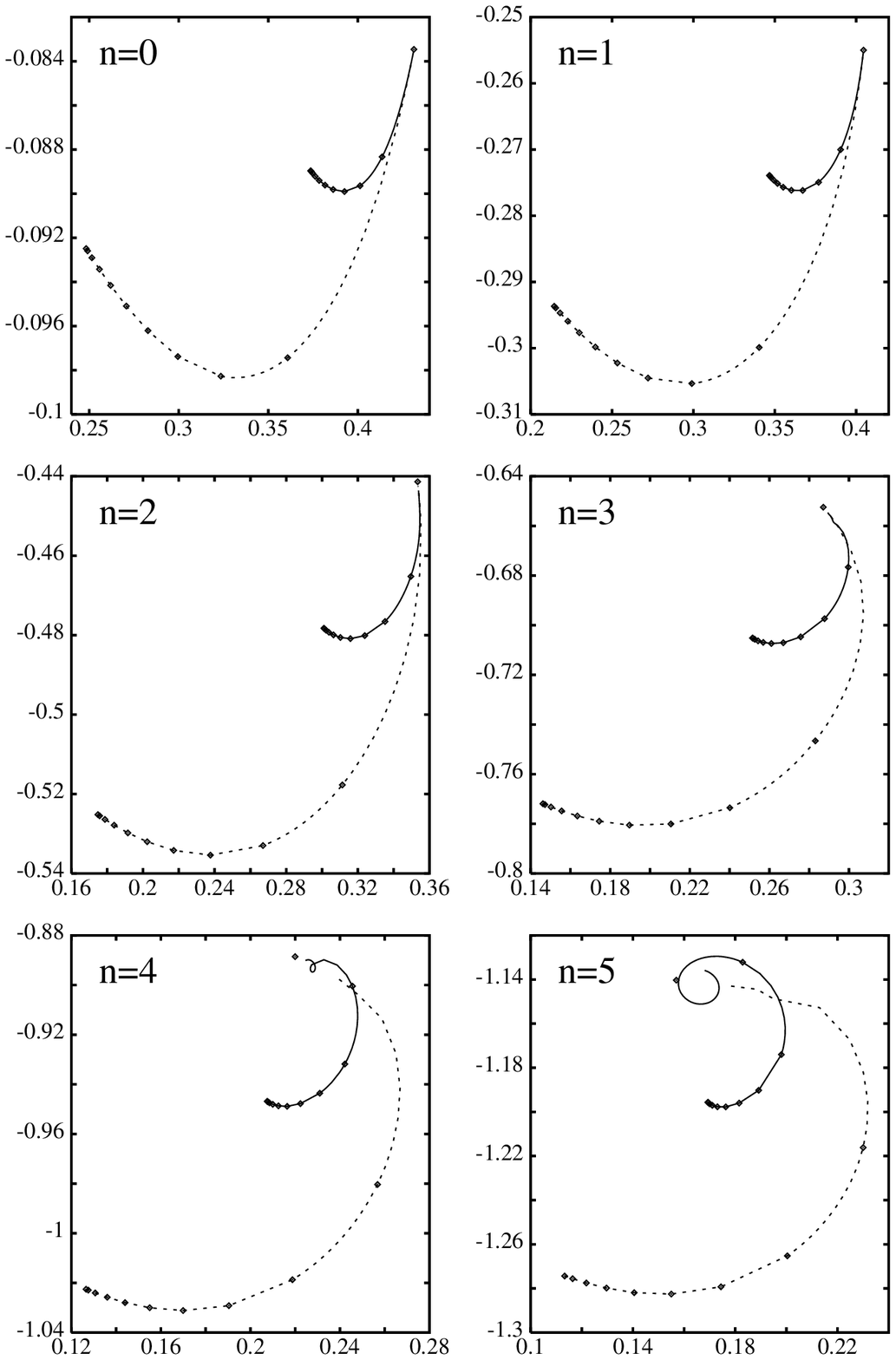}}
\caption{ The behaviour of the  first six quasinormal mode frequencies
(we show $\mbox{Im } \omega M$ as a function of $\mbox{Re } \omega M$)
for a Reissner-Nordstr\"om black hole as the charge is increased. The
solid curves correspond to $\ell=2$ and $\Psi_2$ (which reduces to pure
gravitational waves in the Schwarzschild limit) and the dashed curves
to $\ell=1$ and $\Psi_1$ (that limits to pure electromagnetic waves).
To show that the mode-frequencies change dramatically for $Q>0.9M$ we
have indicated the charge of the black hole by diamonds [at increments
in $Q$ of $0.1M$]. The frequencies generally move anti-clockwise in the
figures as the charge is increased.}
\label{fig1}\end{figure}

For the first few modes the behaviour of the mode-frequencies is
readily described. The damping rate typically reaches a maximum for
$Q/M\approx 0.7-0.8$, and the oscillation frequency generally increases
with $Q$. This is clear from the results for $n=0-3$ in
Fig.~\ref{fig1}. Moreover, this behaviour agrees with the
understanding from the WKB approximation \cite{kokkotas}.  It is
relatively straightforward to verify that, when the lowest order of
approximation is used,  the WKB formulae suggest the following
approximate behaviour (for the slowest damped mode)
\begin{equation}
\mbox{Re }\omega \approx \left( \ell + {1\over 2} \right) \left[ {M\over r_0^3} 
-{Q^2 \over r_0^4} \right]^{1/2} \ ,
\end{equation}
\begin{equation}
\mbox{Im }\omega \approx -{1\over 2} \left[ {M\over r_0^3} 
-{Q^2 \over r_0^4} \right]^{1/2} \left[ {3M\over r_0} 
-{4Q^2 \over r_0^2} \right]^{1/2} \ ,
\end{equation}
in the limit $\ell\gg 1$. Here we have defined $r_0$ as the position
where the black-hole potential attains its maximum value. This means
that $2 r_0 \approx 3M + \sqrt{9M^2 - 8Q^2}$. That is, $r_0$ corresponds to
the position of the unstable, circular photon orbit in the
Reissner-Nordstr\"om spacetime.

To understand better the physics that lead to this behaviour we can use
an argument that is due to Goebel \cite{goebel}:  Consider a congruence
of null rays circling the black hole in the unstable photon orbit. To circle
the black hole would require a coordinate time
\begin{equation}
\Delta t = 2\pi r_0 \left( 1 - {2M\over r_0} + {Q^2 \over r_0^2} \right)^{-1/2} \ .
\end{equation}
The fundamental mode
frequency then follows (if the beam contains $\ell$ cycles) from 
\begin{equation}
\omega \approx {2\pi \ell \over \Delta t}  \ . 
\end{equation}
In a similar way, one can infer the damping rate of the quasinormal
mode
 from the decay rate of the congruence if the null orbit is slightly
perturbed \cite{mashhoon}.
From this information it is easy to convince oneself that the
oscillation
frequency of the modes should increase as $Q$ increases. The results
for the first few modes in Fig.~\ref{fig1} agree nicely with this
description.

A remarkable fact, that is notable in Fig.~\ref{fig1}, is that the
dipole frequencies for $\Psi_1$ approach the quadrupole frequencies for
$\Psi_2$ as the black hole becomes extreme. This effect was first noted by
Onozawa {\em et al.} \cite{onozawa} (for a study of the spin-3/2 case, see \cite{onozawa2}).  They also showed that this
surprising phenomenon arises because the corresponding two effective
potentials are related.  That is, the effective potential  for $\ell$
and $\Psi_1$ is related to that for $\ell+1$ and $\Psi_2$. Specifically, 
$\Psi_1(\ell)$ corresponds to $\Psi_2( -\ell-1)$.
As is easily verified this is true also for the polar potential, but in that case the relation does not take the simple form of
Eq.~(C3) in \cite{onozawa}.


\subsection{Highly damped modes}

\begin{figure}
\centerline{\epsfysize=450pt \epsfbox{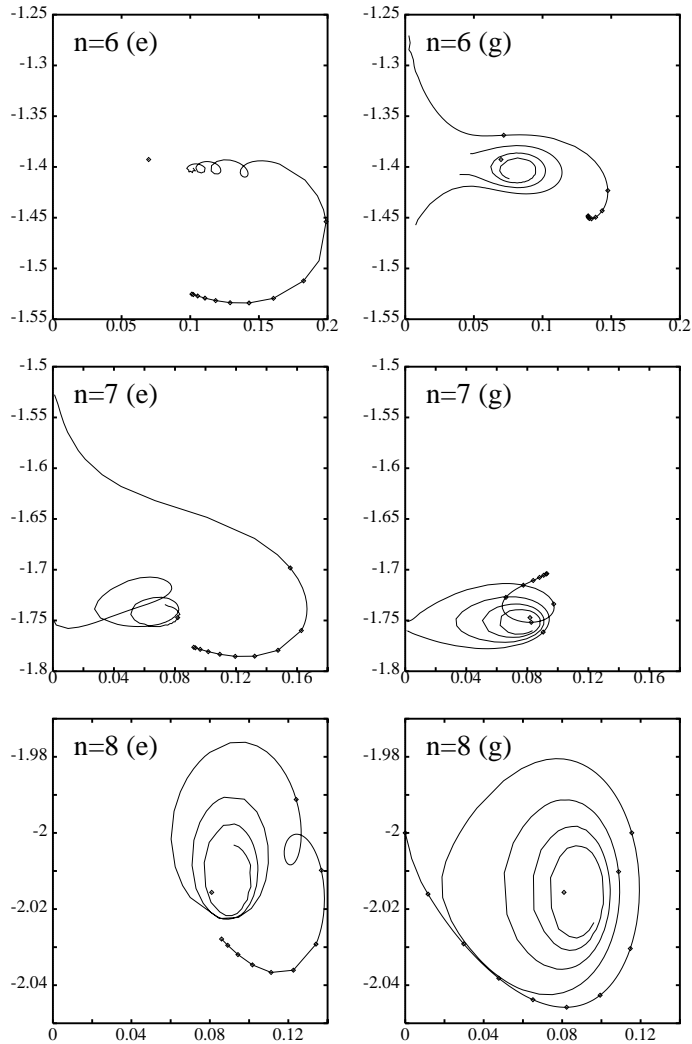}}
\caption{ The behaviour of the  rapidly damped modes (we show $\mbox{Im
} \omega M$ as a function of $\mbox{Re } \omega M$)  of a
Reissner-Nordstr\"om black hole as the charge is increased. The right
frames (labelled `g') correspond to $\ell=2$ and $\Psi_2$ (which
reduces to pure gravitational waves in the Schwarzschild limit) and the
left frames (labelled `e') are for
 $\ell=1$ and $\Psi_1$ (that limits to pure electromagnetic waves). To
 show that the mode-frequencies change dramatically for $Q>0.9M$ we
 have indicated the charge of the black hole by diamonds [at increments
 in $Q$ of $0.1M$]. The frequencies generally move anti-clockwise in
 the figures as the charge is increased.}\label{fig2}\end{figure}

While the behaviour for the first few modes have been studied in detail
before \cite{kokkotas,leaver,andersson} there have been no studies of
the highly damped modes for Reissner-Nordstr\"om black holes. Thus, the
results for $n=5-8$ in Figs.~\ref{fig1} and \ref{fig2} are new. They
are also truly remarkable. While the behaviour is rather simple for the
first modes ($n=0-3$), the rapidly damped modes show many strange
features. Our calculations have discovered a small zoo of 
seemingly different species. Let us discuss them 
separately:

\begin{itemize}
\item The first surprising feature occurs for $n=4$  (see
Figure~\ref{fig1}) at a charge $Q\approx 0.938M $: The $\Psi_2$ mode goes
through a tiny loop that closes at $Q\approx 0.962M$. For the following mode ($n=5$) this loop has
grown. Several similar loops are obvious in the results for $n=6$ and
$\Psi_1$, and when one continues up the spectrum one finds that these
loops are a common feature of many modes.  One interesting aspect of
this result is that  black holes with different charge
may share a specific mode-frequency.

\item For a few modes the mode-frequencies approach the negative
$\mbox{Im }\omega$ axis as the charge varies. Then our numerical
methods \cite{leaver,andersson} become unreliable. This is the reason
for the gaps in the data for $n=6$ for $\Psi_2$ and $n=7$ for both
$\Psi_1$ and $\Psi_2$.  If these modes cross the axis or not is
impossible to say given the available techniques. If they do it would
be very surprising, but given the data in our figures it does not seem
impossible.  One added peculiarity is that it seems as if some modes are ``multi-valued''. That is, for  $n=6$ and $\Psi_2$
we find a mode both in the upper branch close to the imaginary axis
and in the lower branch for a small range of the black-hole charge
($0.8985 \leq Q/M \leq 0.9105 $). A similar behaviour can be observed for 
$n=7$ and $\Psi_1$ when
$0.9330 \leq Q/M \leq 0.9425 $.
At first this result seems nonsensical and obviously
wrong, but it is confirmed by both the continued fraction method and
the numerical integration scheme. This could indicate that it should be
taken seriously.

\item Our study also sheds some light on an issue that has been debated
for Schwarzschild black holes for some time. Is there a quasinormal
mode \underline{on} the imaginary axis at $\omega M = -2i$? Some
methods indicate that there should be such a mode, and that it
corresponds to $n=8$. Other methods say that no such mode exists. To
resolve this issue is difficult, basically because none of
the proposed methods is reliable close to the axis. An added
complication is that the case  $\omega M = -2i$ is a very special one.
That frequency corresponds to a so-called algebraically special
perturbation and, as was shown by Chandrasekhar \cite{cspec}, one can
find analytic solutions to the corresponding perturbation equations.
Although one can convince oneself that Chandrasekhar's special
solutions do not satisfy the quasinormal-mode boundary conditions, the
issue is not resolved. To prove the existence of a quasinormal mode one
must study the analytic properties close to the mode-frequency, and
this turns out to be  difficult in this specific case. Anyway, once
the black hole acquires some small charge it is clear that a mode
exists ($n=8$ for $\Psi_2$ in Fig.~\ref{fig2}). This mode approaches
the suggested value $\omega M = -2i$ as $Q\to 0$. But still this does
not prove the existence of a mode for  the Schwarzschild black hole. It
is plausible that the mode in Fig.~\ref{fig2} and its symmetric
counterpart in the left half of the complex $\omega$-plane will
coalesce at $Q=0$, and perhaps the two modes that exist for the charged
case then cancel each other in some way. It is also worth mentioning here
that there are algebraically special solutions also for the
Reissner-Nordstr\"om black hole.  For $\Psi_{1,2}$ these correspond to
frequencies \cite{cspec}
\begin{equation}
\omega_{1,2}^s = -{i\over 2}\ell(\ell-1)(\ell+1)(\ell+2) \left[
3M\mp\sqrt{9M^2+4Q^2(\ell-1)(\ell+2)} \right]^{-1} \ .
\end{equation}
In general we cannot see any correlation between these 
frequencies and the
quasinormal modes of the charged black hole. 

\item For  higher overtones than those displayed in
Fig.~\ref{fig2} the behaviour tends to be qualitatively similar to the presented ones. The mode-frequency spirals into the value for the
extreme black hole. The spirals get tighter for more rapidly damped
modes.
\end{itemize}

\begin{figure}
\centerline{\epsfysize=200pt \epsfbox{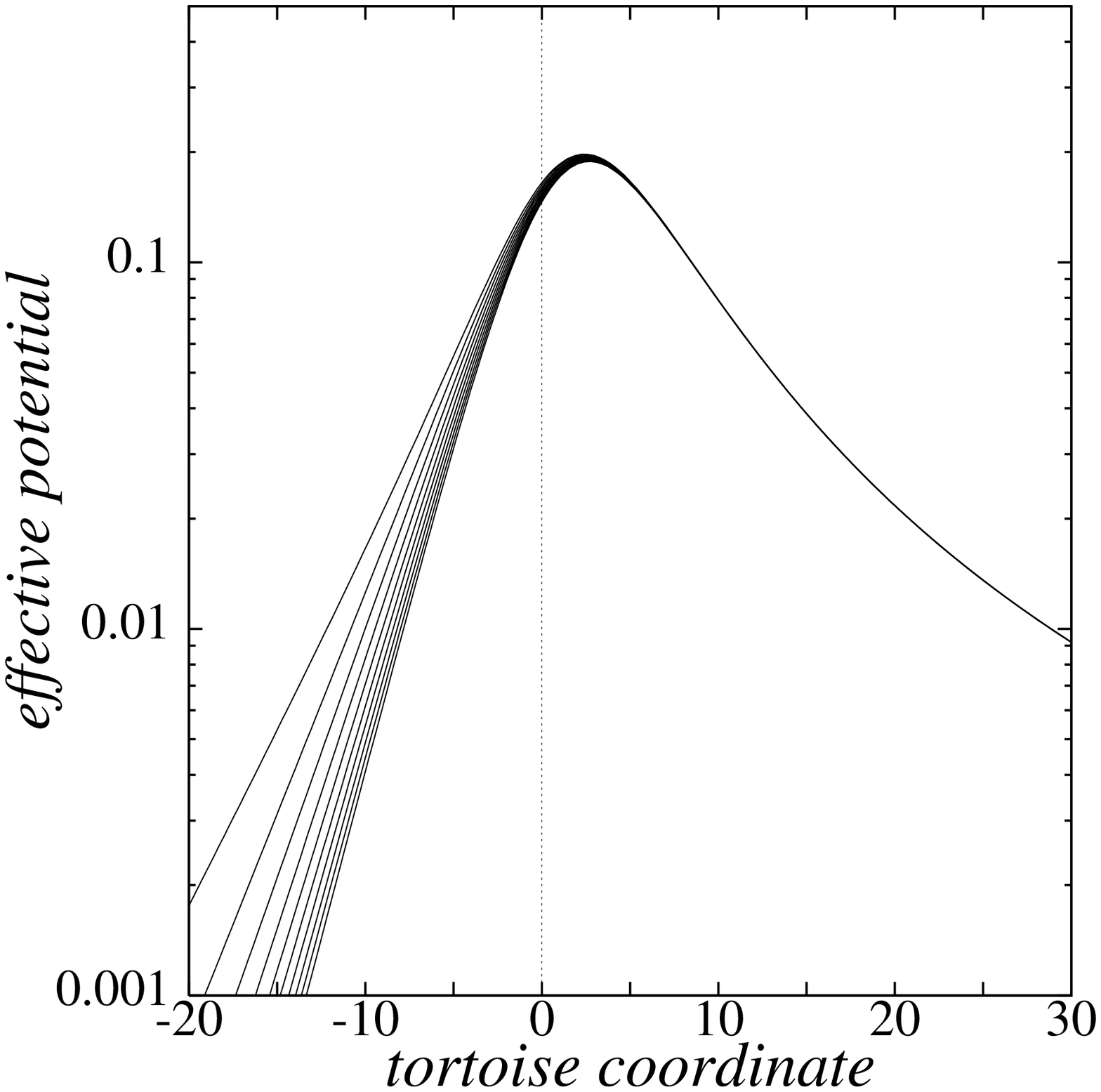}} \caption{ The
effective potential for $\ell=2$ and  $\Psi_2$ (that corresponds to
gravitational perturbations in the Schwarzschild limit) is shown for
$Q/M=0.9-1.0$. The potential falls off slower towards
$r_\ast=-\infty$ as $Q$ increases. The behaviour is similar 
for other values of
$\ell$ and also for the potential that governs $\Psi_1$.}
\label{fig3}\end{figure}

How are we to understand these results? Well, at present it is very
hard to make much sense out of the data presented in Fig.~\ref{fig2}.
The obvious answer is that the change in the mode-frequencies occurs
because the effective potential changes as the charge increases. For the
slowly damped modes it is straightforward to show that the oscillation
frequency depends on the height of the peak of the potential, while the
damping rate is related to the second derivative of $V$ at the peak
\cite{kokkotas}. But it is much harder to draw similar conclusions for
the rapidly damped modes.  Present methods can reliably calculate the
mode-frequencies (and also account for the excitation of the modes in a
dynamical process \cite{leaverprd,andersson95,init}), but we have no
clear understanding of the relation between the highly damped modes and
the details of the effective potential. On the other hand, we have seen
that the mode-frequencies change dramatically as the charge increases
from $Q\approx 0.9M$. If one studies the corresponding change in the
effective potential, cf. Fig.~\ref{fig3}, one can infer that as one
approaches the extreme black hole i) the potential does not change much
for $r\gg r_0$, ii) the peak of the potential 
stays almost constant, and iii) the
main change in the potential is the fall-off rate as
$r_\ast\to-\infty$. This suggests the possibility that the high
overtones of the black hole are in some way related to the effective
potential close to the horizon.  This issue will have to be studied
further in the future.

 
\section{Concluding remarks}

In this paper we have presented the results of a detailed study of the
quasinormal modes of a Reissner-Nordstr\"om black hole. Our work
complements previous studies in several ways. For the first time we
have considered i) the rapidly damped modes, and ii) how the
mode-frequencies approach the anticipated value for the extremely
charged black hole.
We have given special attention to the mode-behaviour near the extreme limit. The
two independent algorithms (from \cite{leaver} and \cite{andersson})
that we used for the calculations provide consistent results even for a
nearly extreme black hole.  These values are also in nice agreement
with the results obtained for the extreme black hole \cite{onozawa}. That is, the quasinormal
modes converge towards the values for the extreme case as $Q\to M$.

We find that while the slowly damped modes can be understood (for
example, by a WKB argument), the behaviour of the rapidly damped modes
as the charge increases is harder to explain. In general, the modes
tend to spiral into the extreme value.  Moreover, there are cases where
the quasinormal modes seem to be double-valued (in cases when the
mode-frequency approaches the imaginary $\omega$-axis).  At one level,
these are just peculiar results without much relevance, but they lead
to several questions. For example; Will the modes of a Kerr black hole
show similar features? This does not seem unlikely, because the 
Kerr problem is in many ways similar to the Reissner-Nordstr\"om one.
Also in the Kerr case will there be two horizons that merge as the
black hole becomes extreme.  Although the slowest damped
modes of the Kerr black hole have been calculated 
\cite{leaverprs}, the rapidly
damped ones have not yet been considered.
Since rotating black holes should be
astrophysically relevant a study of the rapidly damped modes for Kerr
could yield physically interesting results.

The present study also leads to questions regarding the quasinormal
modes themselves: Which features of the effective potential govern the
behaviour of the rapidly damped modes? The present evidence (cf.
Fig.~\ref{fig3}) seems to indicate that the fall-off rate towards the
event horizon plays an important role. But more work is required to
answer the question properly. At present we have to conclude that we
still do not have an intuitive understanding of the origin of the
rapidly damped black-hole modes.

A final question regards the relevance of the algebraically special modes.
In general, our results do not indicate any relation between the
algebraically special mode (that has a purely imaginary frequency) and
the quasinormal modes for a charged black hole. But it is well-known that
the algebraically special mode coalesces with the ninth gravitational
quasinormal mode in the Schwarzschild limit.  Is there a profound reason for this or is it just a
coincidence?  To answer this question is difficult because all present methods
for quasinormal modes break down in the region close to the imaginary frequency axis.  On the other hand, it is known that the algebraically
special modes of the Kerr black hole move away from the imaginary
axis. Hence, it seems likely that a study of the Kerr problem may prove illuminating also for this issue.

Several questions concerning quasinormal modes thus remain to be
resolved.  The present work has contributed a few new pieces
to the puzzle, but many other pieces need to be added before the picture
becomes completely clear. Still,  the present results should
help us get a better understanding not only of the Reissner-Nordstr\"om
black hole, but also of the quasinormal-mode phenomenon in general.

\section*{Acknowledgements}

HO would like to thank A.~Hosoya and H.~Ishihara for continuous
encouragement.  The work of NA is supported by NSF (Grant no PHY
92-2290) and NASA (grant no NAGW 3874). The work of HO is supported in
part by JSPS Research Fellowships for Young Scientists and Scientific
Research Fund of the Ministry of Education.


\newpage

\newpage

\newpage

\begin{references}

\bibitem{vishu} C.V. Vishveshwara  {\em Nature} {\bf 227}, 936 (1970).

\bibitem{press} W.H. Press {\em Ap. J} {\bf 170}, L105 (1971).

\bibitem{nbook} N. Andersson, {\em Black-hole perturbations},
Chapter 4 in {\em Physics of black holes} by I.D. Novikov and V. Frolov
(Kluwer Academic Publishers 1996).


\bibitem{ligo} K.S. Thorne pp. 67--82 in {\it Relativistic
Cosmology}, Proceedings of the Eighth Nishinomiya-Yukawa Memorial
Symposium, Ed: M. Sasaki (Universal Academy Press, Kyoto, 1994).

\bibitem{zerilli} F.J. Zerilli  {\em Phys. Rev. D}  {\bf 9}, 860 (1974).

\bibitem{moncrief} V. Moncrief {\em Phys. Rev. D}  {\bf 9}, 2707 (1974);
  {\bf 10}, 1057 (1974); {\bf 12}, 1526 (1975).

\bibitem{johnston} M. Johnston, R. Ruffini and F. Zerilli {\em Phys. Rev. Lett.}
{\bf 31} 1317 (1973).

\bibitem{olson} D.W. Olson and W.G. Unruh {\em Phys. Rev. Lett} {\bf 33} 1116 (1974).

\bibitem{gunter} D.C. Gunter {\em Phil. Trans. R. Soc. London } {\bf A 299} 37 (1980).

\bibitem{kokkotas} K.D. Kokkotas  and B.F. Schutz  {\em Phys. Rev. D } {\bf 37} 3378 (1988).


\bibitem{leaver} E.W. Leaver {\em Phys. Rev. D} {\bf 41} 2986 (1990).

\bibitem{andersson} N. Andersson {\em Proc. R. Soc. London } {\bf A 442}, 427 (1993).


\bibitem{onozawa} H. Onozawa, T. Mishima, T. Okamura and H. Ishihara
{\em Phys. Rev. D} {\bf 53 } 7033 (1996).


\bibitem{onozawa2} H. Onozawa, T. Okamura, T. Mishima and H. Ishihara {\em Perturbing Supersymmetric Black Hole} Preprint no. TIT/HEP-336/COSMO-74 (1996).
  


\bibitem{leaverprs} E.W. Leaver {\em Proc. R. Soc. London} {\bf A 402} 285 (1985).


\bibitem{cbook}  S. Chandrasekhar {\em The mathematical theory of black holes} 
(Oxford University Press 1983).


\bibitem{anderssonprs}  N. Andersson {\em Proc. R. Soc. London} {\bf A 439} 47 (1992).

\bibitem{goebel} C.J. Goebel {\em Ap. J.} {\bf 172} L95 (1972).

\bibitem{mashhoon} B. Mashhoon {\em Phys. Rev. D} {\bf 31}, 290 (1985).

\bibitem{leaverprd}  E.W. Leaver,  {\em Phys. Rev. D} {\bf 34} 384 (1986).

\bibitem{andersson95} N. Andersson, {\em Phys. Rev. D} {\bf 51} 353 (1995).


\bibitem{init} N. Andersson {\em Evolving wave-fields in a black hole geometry} Preprint no. WUGRAV96-6 (1996).

\bibitem{cspec} S. Chandrasekhar {\em Proc.  R.  Soc. London}  {\bf A  392}, 1 (1984).

\end{references}
\end{document}